\def\year{2021}\relax
\documentclass[letterpaper]{article} 
\usepackage{aaai21}  
\usepackage{times}  
\usepackage{helvet} 
\usepackage{courier}  
\usepackage[hyphens]{url}  
\usepackage{graphicx} 
\usepackage{natbib}  
\usepackage{caption} 
\nocopyright
\newcommand{\mhide}[1]{}

\usepackage{amsmath}
\usepackage{amssymb}
\usepackage{textcase}
\usepackage{soul}
\usepackage{indentfirst}

\usepackage{color}
\usepackage{graphicx} 
\usepackage{fancyvrb}
\usepackage{enumerate}
\usepackage{booktabs}

\usepackage{bbm}
\usepackage{amsfonts}
\usepackage{amsthm}
\usepackage{epsfig}
\usepackage{hyperref}
\usepackage{parskip}
\usepackage{tablefootnote}

\newcommand{\mat}[1]{\boldsymbol{#1}}
\renewcommand{\vec}[1]{\boldsymbol{\mathrm{#1}}}

\newcommand{\RR}{\mathbb{R}}

\providecommand{\mX}{\ensuremath{\mat{X}}}

\providecommand{\vx}{\ensuremath{\vec{x}}}

\theoremstyle{definition}
\newtheorem{definition}{Definition}

\title{Revisiting graph neural networks and distance encoding from a practical view}
\author {
    Haoteng Yin,\textsuperscript{\rm 1}
    Yanbang Wang, \textsuperscript{\rm 2}
    Pan Li \textsuperscript{\rm 1} \\
}
\affiliations {
    \textsuperscript{\rm 1} Department of Computer Science, Purdue University \\
    \textsuperscript{\rm 2} Department of Computer Science, Stanford University \\
    \{yinht, panli\}@purdue.edu, ywangdr@cs.stanford.edu
}

\begin{document}

\maketitle

\begin{abstract}
Graph neural networks (GNNs) are widely used in the applications based on graph structured data, such as node classification and link prediction. However, GNNs are often used as a black-box tool and rarely get in-depth investigated regarding whether they fit certain applications that may have various properties. A recently proposed technique distance encoding (DE)~\cite{li2020distance} magically makes GNNs work well in many applications, including node classification and link prediction. The theory provided in ~\cite{li2020distance} supports DE by proving that DE improves the representation power of GNNs. However, it is not obvious how the theory assists the applications accordingly. Here, we revisit GNNs and DE from a more practical point of view. We want to explain how DE makes GNNs fit for node classification and link prediction. Specifically, for link prediction, DE can be viewed as a way to establish correlations between a pair of node representations. For node classification, the problem becomes more complicated as different  classification tasks may hold node labels that indicate different physical meanings. We focus on the most widely-considered node classification scenarios and categorize the node labels into two types, community type and structure type, and then analyze different mechanisms that GNNs adopt to predict these two types of labels. We also run extensive experiments \footnote{The source code is available at \url{https://github.com/VeritasYin/DEGNN_node_classification}} to compare eight different configurations of GNNs paired with DE to predict node labels over eight real-world graphs. The results demonstrate the uniform effectiveness of DE to predict structure-type labels. Lastly, we reach three pieces of conclusions on how to use GNNs and DE properly in tasks of node classification.
\end{abstract}
\section{Introduction}

\begin{figure}
    \centering
    \includegraphics[trim={0.7cm 8.8cm 0.8cm 4.7cm},clip,width=0.5\textwidth]{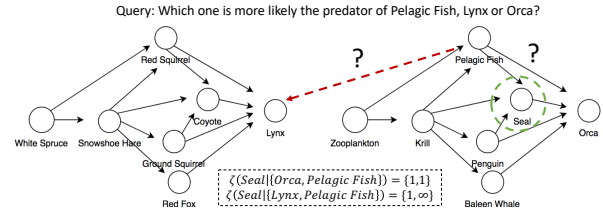}
    \caption{\small Distance encoding helps link prediction in foodweb. These are two isomorphic components in a foodweb graph \cite{srinivasan2019equivalence}.}
    \label{fig:foodweb}
\end{figure}

Graph representation learning has attracted unprecedented attention recently due to the ubiquitous graph-structured data~\cite{hamilton2017representation}. There are a great diversity of graph-related applications ranging from node classification and link (or subgraph) prediction to the entire graph classification. Graph neural networks (GNNs) are a powerful tool for graph representation learning~\cite{scarselli2008graph,battaglia2018relational}. To represent a structure of interest, GNNs in general follow two steps: (1) Learn the representations of nodes; (2) Read out the representations of a group of nodes of interest to make predictions. For example, in link prediction tasks, the representations of two nodes in the link are readout to represent this link~\cite{kipf2016variational,hamilton2017inductive}. This type of design seems to be reasonable. However, there is a severe issue: Node representations learnt by GNNs only reflect the contextual structure around the corresponding nodes; The correlation between two node representations cannot be sufficiently established. Here, correlation between two nodes means a measure regarding the common contextual structure of two nodes, e.g., their distance. Due to the absence of correlation, GNNs may not achieve desired accuracy of the inference that depends on correlations between multiple nodes. An illustrative example of this issue, given by \citeauthor{srinivasan2019equivalence}, is shown in Fig.~\ref{fig:foodweb}. Two nodes in a food web, Lynx and Orca, obtain the same node representations as they have the isomorphic contextual structures. However, if one wants GNNs to tell whether Lynx or Orca is more likely to be the predator of Pelagic Fish (a link prediction task), GNNs cannot make a correct prediction, since node representations of both Lynx and Orca are the same. Only if we associate each node with informative node attributes that reflect its identity, GNNs can distinguish Lynx and Orca in this case.

\begin{figure}[t]
    \centering
    \includegraphics[trim={1.7cm 4.1cm 2cm 3.5cm},clip,height=0.25\textwidth]{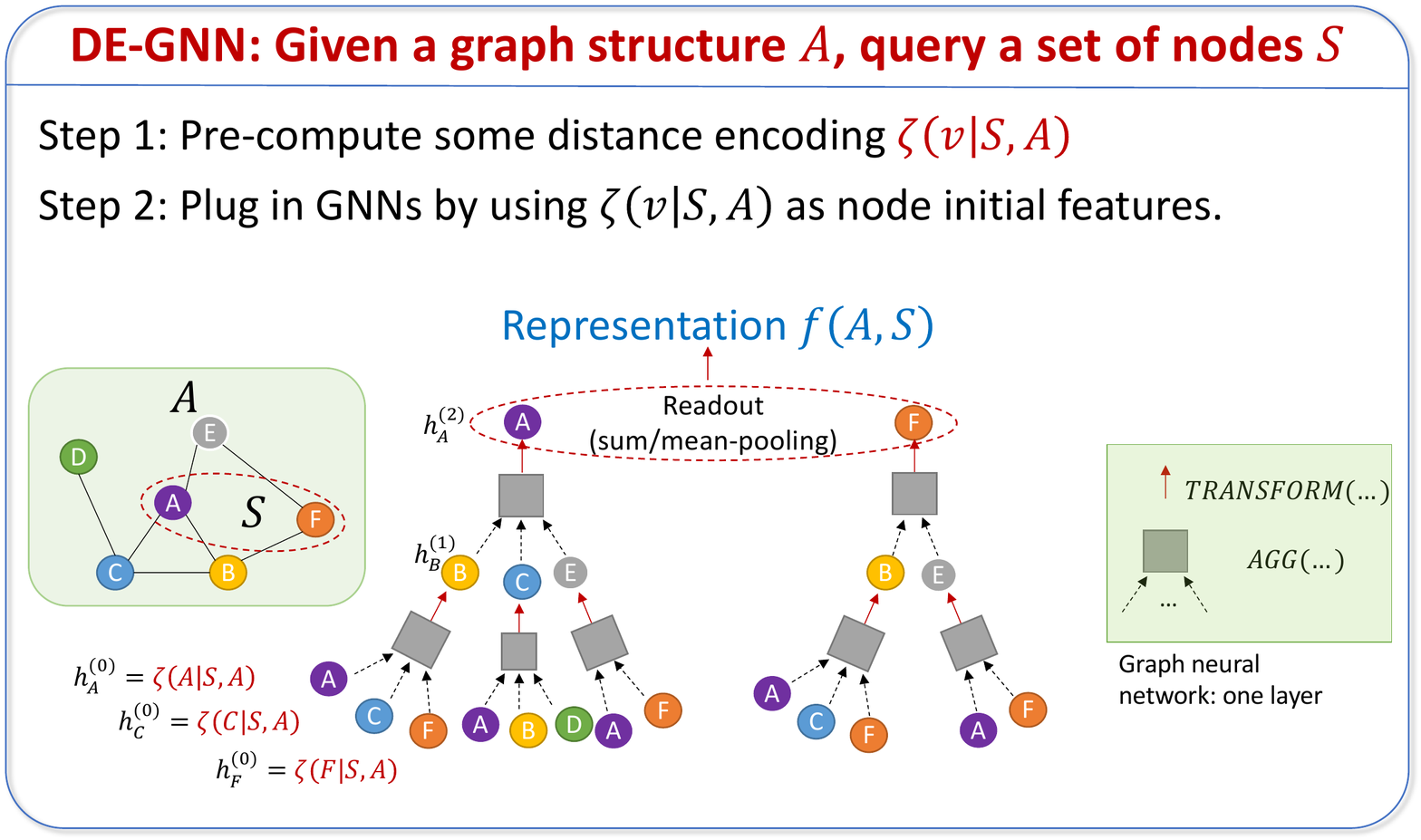}
    \includegraphics[trim={2.7cm 6cm 2.5cm 4.4cm},clip,height=0.20\textwidth]{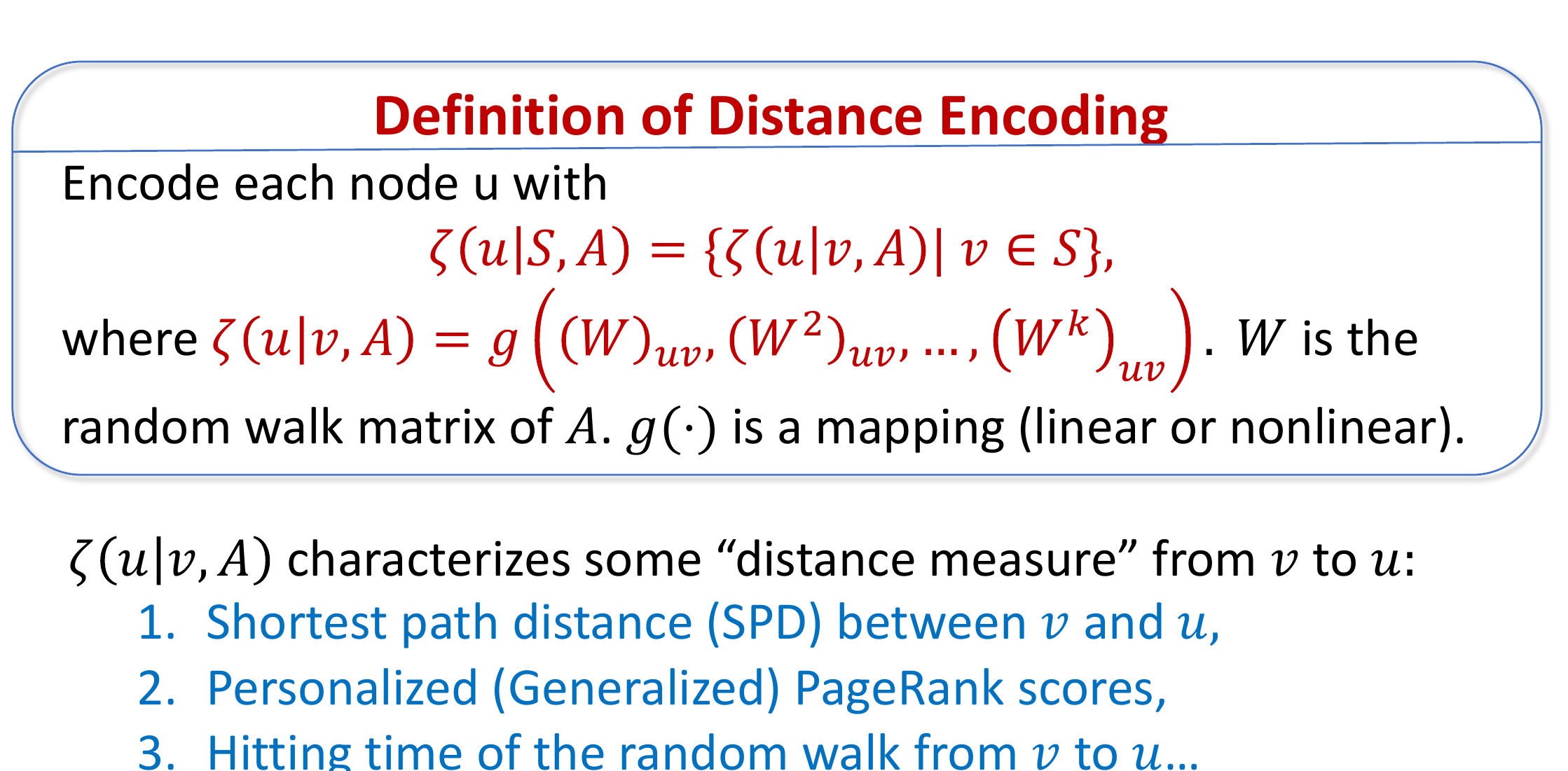}
    \caption{\small The definition of distance encoding \cite{li2020distance}. It can be plugged in GNNs to represent a set of nodes $S$.} 
    \label{fig:DE-GNN}
\end{figure}

Inspired by the empirically successful work~\cite{zhang2018link}, \citeauthor{li2020distance} recently proposed distance encoding techniques (DE) to solve this issue \cite{li2020distance}. They paired the general GNN framework with DE, and obtained the model DE-GNN (see Fig.~\ref{fig:DE-GNN}). DE-GNN assigns every node in the graph with a set of correlated distances to establish the correlation between the representations of a group of nodes of interest. In the case of Fig.~\ref{fig:foodweb}, the node Seal is 1-hop away from Pelagic Fish, which has different distances from it to Orca and Lynx (1-hop and infinite-hop, respectively). Such different distances empower DE-GNN to distinguish relations between these two types of node pairs, (Pelagic Fish, Orca) v.s. (Pelagic Fish, Lynx), and thus leads to a successful prediction. Note that, this procedure is naturally generalized to any higher-order structure (containing more than one node) predictions.

\begin{figure}[htp]
\label{fig:roc}
\begin{center}
\includegraphics[trim={0cm 0cm 0cm 0cm},clip, height=0.5\linewidth]{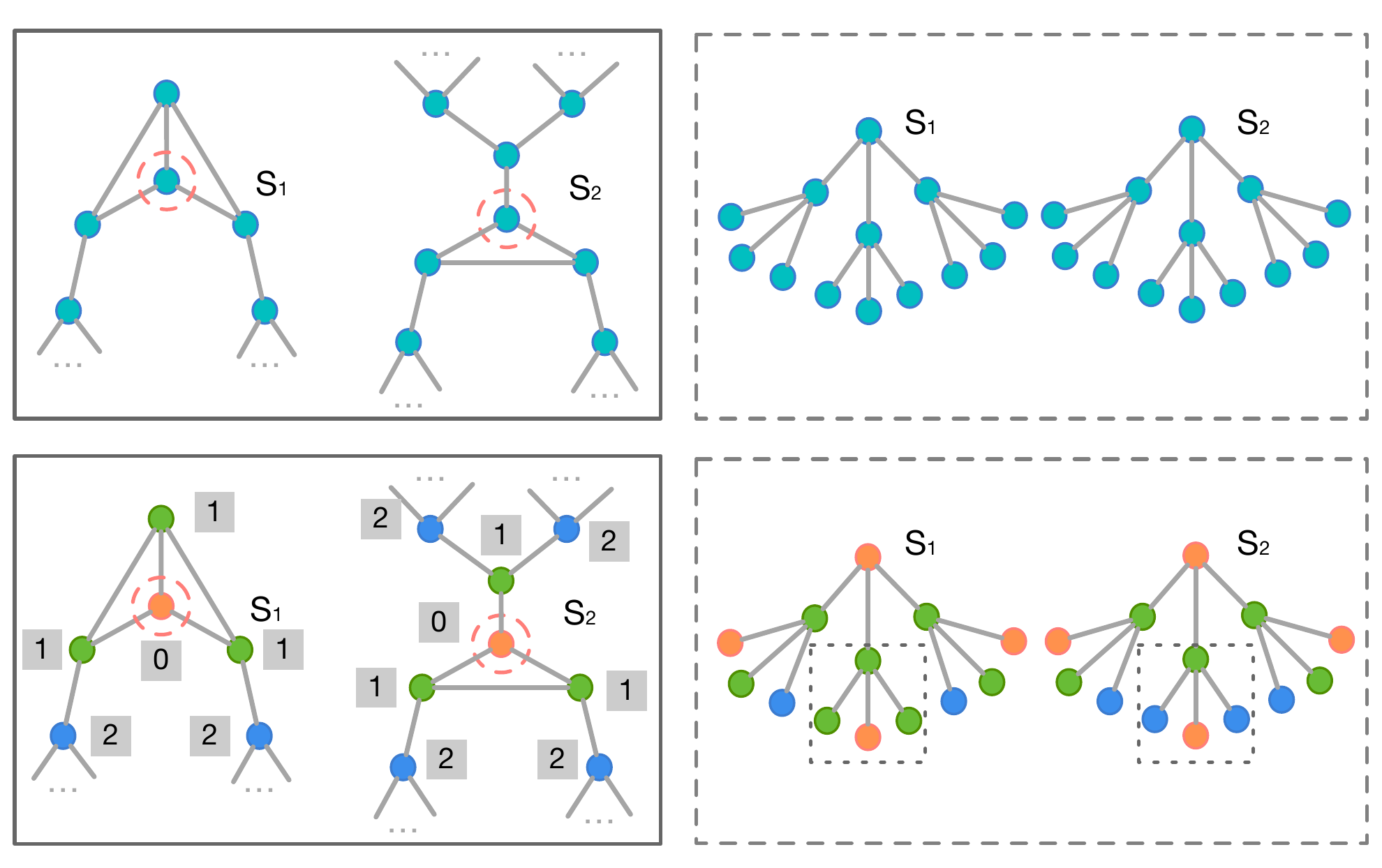}
\end{center}
   \caption{\small Distance encoding helps distinguish contextual structures around nodes: GNNs cannot distinguish the two nodes $S_1$ and $S_2$ with two layers (top left), although they have different contextual structures within their 2-hop neighbors. This is because their computation subtrees rooted at these two nodes are the same (top right). DE-GNN can distinguish $S_1$ and $S_2$ within 2 layers as the computation trees rooted as these two nodes become different by plugging in distance encoding (bottom right).}
\label{fig:node-classification}
\end{figure}

Although the success of DE for the higher-order structure prediction is clear and significant, how DE helps GNNs for node classification is still not clear. This corresponds to the case when the set $S$ in Fig.~\ref{fig:DE-GNN} stands for a single node that is to be classified. \citeauthor{li2020distance} theoretically proves that the representation power of DE-GNN is better than traditional GNNs (see an example in Fig.~\ref{fig:node-classification})\cite{li2020distance}. However, whether GNNs need more representation power of structures to perform good node classification remains unknown. Note that, the representation power is only related to how well GNNs can fit the training set, it does not necessarily imply good testing performance. In other words, if a more representative GNN memorizes any information that is irrelevant to its current task, the testing performance may get deteriorated. Moreover, there could be different formulations of node classification tasks. We need to provide a better understanding of how and where DE fits these real-world tasks of node classification. Otherwise, practitioners may find it challenging to properly use DE and GNNs in their applications, as different combinations of them would be suitable for different settings.

\begin{figure}[t]
    \centering
    \includegraphics[trim={7.3cm 7.2cm 6.1cm 6cm},clip,height=0.245\textwidth]{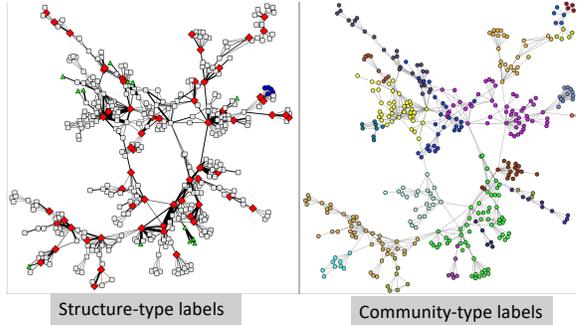}
    \caption{\small A co-authorship network with nodes colored by structure-type labels (left) and community-type labels (right) provided by \cite{henderson2012rolx}.}
    \label{fig:coauthor}
\end{figure}

Previous GNN-based node classification tasks mostly focus on predicting the labels that reflect the community that one node belongs to~\cite{kipf2016semi,velivckovic2017graph,hamilton2017inductive}. Later, we term these labels as \emph{community-type labels} (C-label). For example, consider a co-authorship network, where each node belongs to a researcher and each edge indicates that two researchers have collaborated before. The C-label of a node could be the area that a researcher works in. Graphs with C-labels are typically homophilic, where nodes with same labels are more likely mutually connected (see Fig.~\ref{fig:coauthor} right). 

DE-GNN~\cite{li2020distance}, however, has not been evaluated to predict C-labels. Instead, it was originally tested on predicting another type of node labels, the structural-roles that in practice correspond to the structural functions of nodes in a  network~\cite{henderson2012rolx,ribeiro2017struc2vec}. We term it as \emph{structure-type labels} (S-label) later. As an example of S-labels, still in the co-authorship network (see Fig.~\ref{fig:coauthor} left), the S-label of a node indicates whether a researcher is a core node (red diamonds) or a peripheral one (green triangles); alternatively, whether he/she works in a group (blue circles) or more independently (grey squares). Another example of S-labels in the real world is the positions of different individuals in an organization, such as managers v.s. employees, faculties v.s. students and so on. S-labels typically are indicated by the contextual structures of nodes. Graphs paired with S-labels are heterophilic, where nodes with same labels are not necessarily adjacent and could be spread all across the graph. Although S-labels have been rarely used to evaluate GNNs until very recently, they are comparably important if not more as opposed to C-labels in many graph-related applications~\cite{ahmed2020role}. There are some other types of node classification as well. But here, we focus our discussion on the above two types as they cover almost all the current applications of node classification. 

Because of the difference between C-labels and S-labels, the mechanisms that GNNs use to predict them are different. For C-labels, as the nodes with the same labels are more likely to be connected (homophilic networks), GNNs work in a way to iteratively smooth the features of a node with those of its neighbors. The final predictions are also smoothed over the graph structure. The obtained smoothness is consistent with the allocation of C-labels over graphs, which makes GNNs work well for C-label prediction\footnote{GNNs may work well for C-label prediction but they do not perform the best. GNNs generally work on a local enclosing subgraph around a node. Hence, GNNs can capture limited topological information to predict community labels. Note that, community labels can only be robustly captured via long-range propagation, as theoretically demonstrated in~\cite{li2019optimizing}. Here, long-range propagation means that features need to be propagated a large number of hops ($\geq$ 10) instead of just $2\sim3$ hops as typical GNNs do~\cite{kipf2016semi,velivckovic2017graph}. This point can actually explain why many models, such as SGC~\cite{wu2019simplifying}, APPNP~\cite{klicpera2018predict} and C\&S~\cite{huang2020combining}, by simply removing non-linear activations and allowing propagating node features far across the graphs, could achieve even better performance in C-label prediction.}. However, the above smoothing procedure is not ideal for S-label prediction, as nodes with the same S-labels are spread across the graph, which induces heterophilic graphs. This setting in general means that the previous mechanism of GNNs cannot be applied. To solve this issue, two ways recently have been proposed to make feature-smoothing procedure of GNNs fit under the heterophilic settings: some work explicitly or implicitly changes the graph structures~\cite{pei2020geom,liu2020non} to make nodes from the same S-labels get connected and smooth their features; the other expects GNNs to adaptively learn and perform as high-pass graph filters that smooth raw features on the nodes which are actually far away~\cite{chien2020joint,zhu2020generalizing}. Both strategies still depend on node features and essentially smooth node raw features, even those features get smoothed may not come from their direct neighbors. For heterophilic networks, DE-GNN adopts a fundamentally different mechanism instead. It learns the representation of the ``purely'' contextual structure around each node and sets it as the node representation. This procedure could be totally independent from raw features, though adding node raw features may or may not help practically. As S-labels are typically indicated by the contextual structures of nodes, DE-GNN could learn a suitable node representation for S-label prediction even without informative node raw features. Its effectiveness on this gets demonstrated by predicting passenger flow volume of different airports, where no node raw features were available in ~\cite{li2020distance}.

Based on previous discussion, we may conclude that standard GNNs were mostly used for C-label prediction which heavily depends on informative raw features, while DE-GNN works for S-label prediction and does not necessarily require raw features. In practice, it can be hard to tell whether C- or S-labels the datasets contain, and whether the raw features are informative for a certain type of labels. For instance, the ``purely'' contextual structure (without node raw features) could provide useful information for C-label prediction as if different communities have different internal subgraph structures. Specifically, researchers who are in the area of computer systems typically work in a group as their research projects have heavy workload in general, while researchers from the theory side tend to work in a more independent way. Then, whether a node belongs to some internal groups (like the blue-circle nodes in Fig.\ref{fig:coauthor} left), could be indicative to determine whether they works in systems or theories. Therefore, DE-GNN has the potential to work better on C-label prediction. For S-label prediction, by definition, DE should always be helpful while the raw features are not necessarily. Therefore, it is inconclusive whether we should combine DE with raw features for prediction of S-labels.

Besides the above analysis, in this work, we provide an extensive study of DE-GNNs for node classification. How DE can be combined with GNNs and with raw features is investigated through eight distinct configurations over eight real-world datasets that contain either C-labels or S-labels. Particularly, we consider whether or not using distance encoding, using and/or propagating raw features, plus the various combinations of these choices. In a high level, we reach three following conclusions while leaving further analytical discussion in the latter sections.
\begin{itemize}
    \item DE always significantly improves the performance of GNNs for S-label predictions. 
    \item Raw features are still informative in those heterophilic networks. However, whether or not allowing them to propagate in GNNs depends on the datasets. 
    \item DE used as node features does not help GNNs much for C-label predictions, although previous empirical work~\cite{klicpera2018predict,klicpera2019diffusion} found that applying distance encoding to change the way of aggregation may help with. In this study, we only consider the former case.
\end{itemize}

\section{GNNs and DE-GNN for Node Classification}
In this section, we summarize the key steps of GNNs and the DE technique ~\cite{li2020distance} for node classification. 
\paragraph{Notation}
Given a graph $\mathcal{G}=(V,E)$ with a node feature matrix $\mX_{V} \in \RR^{n \times d}$, where $V$ is the set of nodes and $E$ is the set of edges. Denote its adjacency matrix as $A \in \{0,1\}^{n \times n}$. Each node $v \in V$ has a feature vector $\vx_v \in \mathbb{R}^d$.  

\emph{Homophily ratio} $\psi\triangleq(|\{(u,v):(u,v)\in E \;\text{and}\; y_u=y_v\}|/ |E|) \in [0,1]$ is the fraction of edges in a graph which connect nodes that have the same class label $y$ (i.e. intra-class edges). Based on the value of $\psi$, graphs can be classified into two types, namely, \emph{homophily} ($\psi\to1$) and \emph{heterophily} (i.e. $\psi\to0$).  It is a quantitative indicator of graph datasets, to determine whether the labels are lean towards the community type or not. Note that, the heterophilic graph does not necessarily mean its labels are the structure type, although these two concepts are often consistent in practice.

\paragraph{Graph Neural Networks (GNNs)} The $l$-th layer of a $L$-layer GNN ($l=1,2,\dots,L)$ can be formulated as
\begin{equation}
\nonumber
\resizebox{0.99\hsize}{!}{%
$r_v^{(l)}=\texttt{AGG}^{(l)}\left(\{h_u^{(l-1)}:u\in \mathcal{N}(v)\}\right), h_v^{(l)}=\texttt{COMBINE}^{(l)}\left(\{h_v^{(l-1)}\},r_v^{(l)}\right)$}
\end{equation}
where $h_v^{(l)}$ is the representation of node $v$ at the $l$-th layer. 

The $\texttt{AGG}^{(l)}(\cdot)$ function determines how to aggregate information from node $v$'s neighborhood $\mathcal{N}(v)$ (such as, max or mean pooling), while the $\texttt{COMBINE}^{(l)}(\cdot)$ function decides how to fuse the representation of node $v$ from previous layer $h_v^{(l-1)}$ and aggregate information from its neighbors $r_v^{(l)}$. 

\paragraph{Problem Setup} We focus on the semi-supervised node classification problem on a simple graph $\mathcal{G}$. Given a training set $V_{\mathcal{T}} \subset V$ with known class labels $y_v \in Y$ for all $v \in V_{\mathcal{T}}$, and a feature vector $\vx_v$ for $v \in V$, we aim to infer the unknown class labels $y_u$ for all $u \in (V - V_{\mathcal{T}})$.

\subsection{Node Structural Features and Distance Encoding}
The summary of graph statistics (e.g. node degrees) are generally used as structural features in many traditional machine learning approaches. Walk-based techniques such as subgraph sampling and random walk, are very popular to be applied for extracting structural information by exploiting node localities of graphs. In the following, we introduce another general class of structure-related features, termed Distance Encoding (DE) \cite{li2020distance}. DE is an approach to encode the distance between a node in the graph and a target node set. In this study, we adopt a simplified version of DE by reducing the size of node set to 1. Suppose we are to classify node $u$, DE between any node in the graph and $u$ is formally defined as follows.

\begin{definition}
Given a target node $u \in V$ for classification and the adjacency matrix $A$, \textit{distance encoding} for any $v\in V$, $\zeta(v|u, A)$, is a mapping based on a collection of landing probabilities of random walks from $u$ to $v$, i.e.,
\begin{equation}
\nonumber
\resizebox{0.99\hsize}{!}{%
$\zeta(v|u,A) = \phi(\ell_{vu}),~\ell_{vu}=[(W)_{vu}, (W^2)_{vu}, \dots, (W^k)_{vu}, \dots],$}
\end{equation}
where $W=AD^{-1}$ is the random walk matrix. $\phi(\cdot)$ can be a learnable or fixed function.
\end{definition}

$\zeta(v|u,A)$ defined above is a general form that covers several graph-related measurements. If we set $\phi(\ell_{vu})$ as the first non-zero position in $\ell_{vu}$, then it gives the \textit{shortest-path-distance (SPD)} from $u$ to $v$, noted as $\zeta_{spd}(v|u)$. $\phi$ could also be a feed-forward neural network. In practice, a finite length of $\ell_{vu}$ is sufficient to encode the contextual structure of a node ($k:2\sim 4$).

Compared to other techniques to represent the structural roles of nodes, e.g. motif/graphlet counting~\cite{ahmed2020role}, DE is computationally efficient. Meanwhile, DE is different from positional node embeddings, such as Node2vec~\cite{grover2016node2vec}, because it captures the relative distance between a node and the target node. Such a distance is independent of the absolute positions of nodes in graphs, which makes sure that DE-based approaches are inductive to predict the labels spread in the sections of graphs that have not been used for training.

\subsection{Utilize Structural Features into Node Classification\label{sec:de}}
In conventional GNNs, the representation $h_v^{(0)}$ for the first layer is initialized by the raw features $\vx_v$. Besides it, structural features $\mX_S \in \RR^{n \times k}$ such as DE and node degree can also be used for initialization by setting initial node features as $\mathcal{X} = (\mX_V \mathbin\Vert \mX_S) \in \RR^{n\times(d+k)}$, where $\mathbin\Vert$ denotes the combining operation between two types of features. In reality, especially for S-label prediction, the raw features from neighbors could be irrelevant while structural features are informative. Thus, we consider GNNs solely propagate structural features first, and then concatenate the last-layer representation of a node (currently only hold ``pure'' structural information) with its raw features to make predictions, which 
eliminates the effects of irrelevant raw features from its neighbors.

Based on a model uses which type(s) of features (node degree, raw and DE features), where to add raw features (the first or the last layer in a model), we design eight types of GNN variants for classification tasks in terms of structural roles. In the subsequent discussion, those variants are indexed from $\textbf{M}_1$ to $\textbf{M}_8$. The detailed configuration of each model is listed in Table \ref{tab:config}. Here, $\textbf{M}_5$ only takes raw features as the initial input without structural features involved, which is equivalent to standard GNNs under a typical classification setting. Similarly, $\textbf{M}_6$ does not apply either raw features or structural features for the propagation, which is essentially degenerated to a multilayer perceptron (MLP).

\begin{table}[t]
\centering
\resizebox{0.48\textwidth}{!}{%
\begin{tabular}{c|clc|l}
\toprule[1pt]
\textbf{Model} & \textbf{DE } & \textbf{Raw Features}
& \textbf{Degree}                      & \textbf{Description} \\ \hline
$\mathbf{M}_1$ & $\checkmark$           & $\checkmark$ (First)  & $\checkmark$      & with all features                                 \\
$\mathbf{M}_2$ & $\checkmark$           & $\checkmark$ (Last)   & $\checkmark$      & with all features                                 \\
$\mathbf{M}_3$ & $\checkmark$           & $\times$           & $\checkmark$      & without raw node features                         \\
$\mathbf{M}_4$ & $\times$          & $\times$            & $\checkmark$      & node degree only                                 \\
$\mathbf{M}_5$ & $\times$          & $\checkmark$ (First)  & $\times$      & raw node features only                      \\
$\mathbf{M}_6$ & $\times$           & $\checkmark$ (Last)   & $\times$      & raw node features only                      \\
$\mathbf{M}_7$ & $\times$          & $\checkmark$ (First)  & $\checkmark$      & without DE features                               \\
$\mathbf{M}_8$ & $\times$           & $\checkmark$ (Last)   & $\checkmark$      & without DE features                               \\
\bottomrule[1pt]
\end{tabular}%
}
\caption{Configuration of GNN variants \label{tab:config}. `$\checkmark$' means this term is used in this model while `$\times$' denotes the opposite. ``First'' or ``Last'' refers to whether the raw features are added to the node representations in the first or last layers.}
\end{table}

\begin{table*}[t]
\resizebox{0.99\textwidth}{!}{%
\begin{tabular}{l|ccc|cc|ccc}
\toprule[1pt]
\textbf{Model} &
  \textbf{Cora} &
  \textbf{Citeseer} &
  \textbf{Pubmed} &
  \textbf{Chameleon} &
  \textbf{Actor} &
  \textbf{Cornell} &
  \textbf{Texas} &
  \textbf{Wisconsin} \\\hline
\textbf{GCN*}             & 85.77 & 73.68 & 88.13 & 28.18 & 26.86 & 52.70 & 52.16 & 45.88 \\
\textbf{Geom-GCN*}             & 85.27 & \textbf{77.99} & \textbf{90.05} & 60.90 & 31.63 & 60.81 & 67.57 & 64.12 \\ \hline
$\textbf{M}_1$-SPD & 85.76$\pm$1.32 & 74.42$\pm$0.88 & 85.74$\pm$0.32 & 65.49$\pm$2.11 & 35.30$\pm$0.96 & 68.38$\pm$7.46 & 79.19$\pm$6.29 & 75.60$\pm$6.74\\ 
$\textbf{M}_2$-SPD & 71.81$\pm$2.18 & 71.61$\pm$2.22 & 85.42$\pm$0.58 & 51.74$\pm$2.67 & 36.26$\pm$1.52 & \textbf{81.35$\pm$3.51} & \textbf{84.05$\pm$3.72} & 83.60$\pm$4.27\\ 
$\textbf{M}_3$-SPD & 39.23$\pm$6.36 & 27.79$\pm$1.31 & 44.65$\pm$0.51 & 48.22$\pm$2.05 & 25.56$\pm$0.75 & 50.27$\pm$7.24 & 53.51$\pm$5.51 & 50.00$\pm$6.19 \\ \hline
$\textbf{M}_1$-RW & 85.63$\pm$1.63 & 73.98$\pm$1.19 & 85.40$\pm$0.42 & \textbf{65.82$\pm$2.30} & 35.12$\pm$0.72 & 65.14$\pm$4.90 & 78.65$\pm$6.10 & 76.60$\pm$4.10\\ 
$\textbf{M}_2$-RW & 71.05$\pm$2.39 & 71.01$\pm$1.43 & 85.76$\pm$0.52 & 52.00$\pm$2.04 & \textbf{36.42$\pm$1.11} & 81.08$\pm$4.83 & 82.97$\pm$5.28 & \textbf{84.00$\pm$4.98}\\  
$\textbf{M}_3$-RW & 37.77$\pm$2.68 & 27.46$\pm$1.88 & 46.19$\pm$0.75 & 39.03$\pm$3.45 & 36.13$\pm$2.03 & 51.89$\pm$3.59 & 53.24$\pm$5.80 & 51.20$\pm$4.02 \\ \hline
$\textbf{M}_4$ & 36.81$\pm$3.91 & 27.41$\pm$1.38 & 44.46$\pm$0.52 & 35.76$\pm$3.99 & 25.53$\pm$0.36 & 51.62$\pm$4.09 & 54.05$\pm$6.04 & 50.60$\pm$3.90\\
\textbf{$\textbf{M}_5$ (SAGE)} & 86.03$\pm$1.94 & 74.24$\pm$1.39 & 86.38$\pm$0.57 & 60.75$\pm$2.01 & 34.38$\pm$0.96 & 69.73$\pm$6.10 & 76.76$\pm$5.82 & 78.60$\pm$3.58 \\
\textbf{$\textbf{M}_6$ (MLP)} & 71.16$\pm$1.93 & 71.32$\pm$1.22 & 85.12$\pm$0.50 & 51.78$\pm$1.83 & 35.76$\pm$1.15 & 80.00$\pm$3.86 & 81.62$\pm$2.65 & 81.00$\pm$4.40 \\
$\textbf{M}_7$ & \textbf{86.42$\pm$1.79} & 73.92$\pm$1.42 & 86.35$\pm$0.55 & 61.82$\pm$1.69 & 34.36$\pm$1.17 & 71.89$\pm$6.64 & 79.46$\pm$4.55 & 78.20$\pm$4.51\\
$\textbf{M}_8$ & 72.99$\pm$1.61 & 71.80$\pm$1.73 & 85.03$\pm$0.76 & 49.67$\pm$1.34 & 35.49$\pm$0.93 & 78.65$\pm$5.33 & 78.65$\pm$6.33 & 83.00$\pm$4.02\\
\bottomrule[1pt]
\end{tabular}
}
\caption{Comparison of models' performance in mean accuracy (\%) $\pm$stdev based on 10 times random tests. Models with asterisk `*' indicate the results are obtained from \cite{pei2020geom}. Best model performance per dataset is marked in bold.\label{tab:res}}
\end{table*}

\section{Experiments} \label{sec:exp}
\subsection{Datasets \& Baselines}
We utilize three types of public datasets for node classification tasks, namely, citation networks, Wikipedia network and WebKB. An overview of characteristics for each dataset is summarized in Table \ref{tab:dataset}.

Citation networks include Cora, Citeseer, and Pubmed \cite{sen2008collective,namata2012query}, which are a standard benchmark for semi-supervised node classification. In these networks, nodes and edges denote for papers and citation relation between them, respectively. Node features are the embeddings of papers, whose labels are C-labels that indicate the research fields of the papers. Citation networks have a strong tendency of homophily with ratios $\psi>0.7$.

Chameleon and Actor are subgraphs of webpages in Wikipedia \cite{rozemberczki2019multi,tang2009social}. Nodes represent for webpages regrading specific topics or actor personals, and edges denote mutual links between pages or co-occurrence in the same page. We use the labels generated by \citeauthor{pei2020geom}: nodes in Chameleon are classified into five categories based on monthly averaged visits. The node labels are S-labels, as the visiting traffic implies whether a node is a hub or not. Nodes in Actors have 5 classes in term of words appeared in the actor’s Wikipedia page, while it is hard to tell whether they are C- or S- labels. Both graphs have low homophily ratios around 0.2.

Cornell, Texas, and Wisconsin are three subset of WebKB dataset, which are the network map of university-wise web-pages collected by CMU. In these networks, each node denotes a web page, which is manually classified into 5 categories: student, staff, faculty, course and project. The labels indicate the structural roles of nodes in terms of of job positions and oriented audiences, and thus are S-labels.

\begin{table}
\resizebox{0.48\textwidth}{!}{%
\begin{tabular}{l|ccc|cc|cccc}
\toprule[1pt]
\textbf{Dataset} &
  \textbf{Cora} &
  \textbf{Cite.} &
  \textbf{Pubm.} &
  \textbf{Cham.} &
  \textbf{Actor} &
  \textbf{Corn.} &
  \textbf{Wisc.} &
  \textbf{Texa.} \\ \hline
 \textbf{Label Type} & C & C & C & S & - & S & S & S \\\hline
\textbf{Ratio $\mathbf{\psi}$}   & 0.81  & 0.74  & 0.8    & 0.23   & 0.22   & 0.3   & 0.21  & 0.11  \\ \hline
\textbf{\# Nodes}    & 2,708 & 3,327 & 19,717 & 2,277  & 7,600  & 183   & 183   & 251   \\
\textbf{\# Edges}    & 5,428 & 4,732 & 44,338 & 36,101 & 33,544 & 295   & 309   & 499   \\
\textbf{\# Features} & 1,433 & 3,703 & 500    & 2,325  & 931    & 1,703 & 1,703 & 1,703 \\
\textbf{\# Class}    & 7     & 6     & 3      & 4      & 4      & 5     & 5     & 5     \\
\bottomrule[1pt]
\end{tabular}%
}
\caption{Datasets statistics \label{tab:dataset}}
\end{table}

For comparison, we select three representative models GCN~\cite{kipf2016semi}, GraphSAGE~\cite{hamilton2017inductive}, and Geom-GCN~\cite{pei2020geom} as baseline models. We also consider MLP with one hidden layer as a graph-agnostic baseline. (Graph)SAGE and MLP correspond to $\mathbf{M}_5$ and $\mathbf{M}_6$, respectively.

\subsection{Benchmark Settings}
For all benchmarks, we follow the same dataset setting as in \cite{pei2020geom} that randomly splits nodes of each class into 60\%, 20\%, and 20\% for training, validation and testing. We perform hyper-parameter searching for all models on the validation set (details are attached in the appendix).

We choose SAGE as the GNN backbone and construct eight GNN variants $\textbf{M}_1\sim\textbf{M}_8$ in Table~\ref{tab:config}. In particular, to evaluate DE variants, we apply both SPD based one-hot vector $\zeta_{spd}$ and the sequence of landing probabilities of random walks. Models with those two types of DEs are labelled as $\mathbf{M}_i$-SPD and $\mathbf{M}_i$-RW for $i\in\{1,2,3\}$, respectively.

\subsection{Results \& Discussion}
We report the results in Table \ref{tab:res}.

\textbf{Graphs with C-labels} On homophilic graphs ($\psi > 0.7$), GNNs with DE provide a similar level of performance against other baselines, even those GCN-based methods that are heavily optimized under a strict homophily assumption. As we can tell from the first three columns of Table \ref{tab:res} and compare $\textbf{M}_1$ with $\textbf{M}_5,\textbf{M}_7$, feeding DE to the model may not obtain performance gains but would not hinder it much. But one may also use DE as controllers of aggregation procedure inside GNNs, termed DEA-GNN in~\cite{li2020distance}. This part is not included in our current study. However, different types of graph diffusion (e.g., personalized-PageRank diffusion, heat-kernel diffusion), as a special type of DE's application on changing the way of propagation, can significantly improve node classification over homophilic graphs \cite{klicpera2019diffusion}.

\textbf{Graphs with S-labels} On heterophilic graphs, we observe consistent improvement on the accuracy of models with structural features (either degree or DE features, or both two). Notably, the best performance for each dataset with S-labels is always one of models with assistance of DE, which applied either SPD or RW. From the outcome of $\textbf{M}_3$, it suggests that node raw features still play a role in classification of S-labels. Furthermore, in this particular case, DE alone might not be sufficient ($\textbf{M}_3$ v.s. $\textbf{M}_4$). However, whether raw fractures $\mX_v$ should be initially propagating through the network or not is inconclusive. As results from both $\textbf{M}_1$/$\textbf{M}_7$ and $\textbf{M}_2$/$\textbf{M}_8$ indicated, it is up to the datasets in practice.

In summary, we can conclude that structural features generalize GNNs on both homophilic and heterophilic graphs for node classification. The above results demonstrate that GNNs with both raw and structural features have the best performance on S-label graphs while maintaining comparable performance on graphs with the type of C-labels. GNNs with DE assisted are a more universal framework for both graph settings, compared to other models that have an implicit assumption of strong homophily for the dataset. In particular, DE is significantly useful to predict S-labels and naturally makes GNNs fit for heterophilic graphs. 

\section{Conclusion}
In this work, we readdressed the issue and the potential of graph neural networks on semi-supervised node classification. Specifically, we investigated how structural features such as distance encoding would work with raw features under this setting, and then conducted an extensively empirical study on their individual and collaborative effectiveness for GNNs in node classification tasks. Our study includes both homophilic and heterophilic graphs with community- or structure-type labels. The experiments show that structural information is able to generalize GNNs on both types of labels and graph settings. It particularly demonstrates the uniform effectiveness of applying distance encoding to assist GNNs on prediction of structure-type labels.

\bibliography{ref.bib}
\newpage
\section{Appendix}
\subsection{Hyper-parameter Search}
For a fair comparison, the size of search space for each model is set to the same, including number of hidden unit, initial learning rate, dropout rate, and weight decay, which is listed below.

\begin{itemize}
    \item \textbf{SAGE}
    \begin{itemize}
        \item \texttt{num\_layer:} \texttt{n} $\in$ \texttt{\{1, 2, 3\}}
        \item \texttt{hidden\_dim:} \texttt{n} $\in$ \texttt{\{32, 64, 128, 256\}}
        \item \texttt{learning rate:} \texttt{lr} $\in$ \texttt{[1e-5, 1e-3]}
        \item \texttt{epochs:} \texttt{200 / 500}
        \item \texttt{patience:} \texttt{50}
        \item \texttt{dropout rate:} \texttt{p} $\in$ \texttt{[0.1, 0.9]}
        \item \texttt{weight decay:} \texttt{l2} $\in$ \texttt{\{1e-6, 1e-5, 5e-5\}}
    \end{itemize}
    \item \textbf{MLP}
    \begin{itemize}
        \item \texttt{num\_layer:} \texttt{n} $\in$ \texttt{\{1, 2\}}
        \item \texttt{hidden\_dim:} \texttt{n} $\in$ \texttt{\{32, 64, 128\}}
        \item \texttt{non-linearity}: \texttt{ReLU}
        \item \texttt{dropout rate:} \texttt{0}
    \end{itemize}
\end{itemize}

For our SAGE benchmark, the number of hidden layers is set to 1, expect for Cora, Citeseer and Actor which would need 2 or 3 layers; the number of hidden unit is 32 (Cora, Pubmed and Chameleon), 64 (Citeseer, WebKB), and 256 (Actor). The final hyper-parameter setting for training: initial learning rate \texttt{lr=1e-4} with early stopping (patience of 50 epochs); weight decay \texttt{l2=1e-6} for Cora, Citeseer and Actor; dropout \texttt{p=0.2} for Citeseer, Actor and WebKB while \texttt{p=0.4} for Cora and Chameleon. We use Adam \cite{kingma2014adam} as the default optimizer. 

Compared to the official PyTorch implementation of DE-GNN, our version introduces the following new features:
\begin{itemize}
    \item the framework is able to repeatedly train instances of models without regenerating graph samples;
    \item the procedure of subgraph extraction and the number of GNN layers are decoupled. The model now can run on subgraphs with arbitrary number of hops, but without the limit of binding the extraction hops with the total number of layer propagation in GNNs;
    \item sparse matrices are utilized to accelerate the computation of DE, especially for random-walk based features.
\end{itemize}
\end{document}